\documentclass[letterpaper, superscriptaddress, twocolumn, showpacs,
nofootinbib]{revtex4}

\usepackage[]{graphicx}
\usepackage{amsmath}
\usepackage{amsfonts}
\usepackage{color}
\usepackage{hyperref}
\usepackage{bm}
\usepackage{graphicx}
\usepackage{amsbsy}
\usepackage{epstopdf}

\def\ket#1{| #1 \rangle}
\def\bra#1{\langle #1 |}
\def\kb#1#2{|#1\rangle\!\langle #2 |}
\def\bk#1#2{\langle #1 | #2 \rangle}

\def\id{{\mathchoice {\rm 1\mskip-4mu l} {\rm 1\mskip-4mu l} {\rm
1\mskip-4.5mu l} {\rm 1\mskip-5mu l}}}
\newcommand\etal{{\em et al.}}

\definecolor{gold}{rgb}{0.85,.66,0}
\definecolor{brown}{rgb}{0.647,0.165,0.165}

 
\date{\today}

\begin{document}

\title{\textcolor{black}{Quantum state cloning in the presence of a closed timelike curve}}

\author{D.~Ahn}\email{dahn@uos.ac.kr}
\affiliation{Centre for Quantum Information Processing, Department of Electrical and Computer Engineering, University of Seoul, Seoul 130-743, Republic of Korea}
\author{C.~R.~Myers}\email{myers@physics.uq.edu.au}
\affiliation{Centre for Engineered Quantum Systems, School of Mathematics and Physics,
The University of Queensland, St. Lucia 4072 QLD Australia}
\author{T.~C.~Ralph}
\affiliation{Centre for  Quantum Computation and Communication Technology, School of Mathematics and Physics,
The University of Queensland, St. Lucia 4072 QLD Australia}
\author{R.~B.~Mann}
\affiliation{Department of Physics and Astronomy, University of Waterloo, Waterloo, Ontario, N2L, 3G1, Canada}

\begin{abstract}
Using the Deutsch approach, we have shown that the no-cloning theorem can be circumvented in the presence of closed timelike curves, allowing the perfect cloning of a quantum state chosen randomly from a finite alphabet of states. Further, we show that a universal cloner can be constructed that when acting on a completely arbitrary qubit state, exceeds the no-cloning bound on fidelity. Since the no cloning theorem has played a central role in the development of quantum information science, it is clear that the existence of closed timelike curves would radically change the rules for quantum information technology. Nevertheless, we show that this type of cloning does not violate no-signalling criteria. 
\end{abstract}

\maketitle


The possible existence of closed timelike curves (CTCs) allowing time travel~\cite{Morris88a, Gott91a} draws attention to fundamental questions about what is physically possible and what is not~\cite{Deutsch91a, Politzer94a, Hartle94a, Cassidy95a, Hawking95a, Bacon04a, Ralph07a, Brun09a, Bennett09a}. An example is the impossibility theorem in quantum mechanics called the ``no cloning theorm''~\cite{Wooters82a, Dieks82a, Barnum96a}, which states that there exists no physical means by which an unknown arbitrary quantum state can be reproduced or copied if chronology is respected. Recently, Brun~\etal~\cite{Brun09a} showed that in the presence of a CTC it was possible to perfectly discriminate a finite set of non-orthogonal quantum states of a two-level system - a qubit. They conjectured that universal, CTC-assisted cloning with fidelity approaching one was possible, at the cost of increasing the available dimensions in ancillary and CTC resources. 

Here we show explicitly that in the presence of closed timelike curves quantum mechanics allows the cloning of an unknown arbitrary quantum state with fidelity that exceeds the no-cloning bound \cite{Buzek96a, Scarani05a}, using a finite dimensional ancillary and CTC resource. One of the original arguments against cloning was that it would allow signalling, i.e. faster than light communication, when applied to an entangled state~\cite{Wooters82a}. We also show that this type of cloning does not violate the no-signalling criteria. Should the ability to manipulate closed timelike curves ever become possible, our research suggests that new possibilities in quantum information technology would emerge, including eavesdropping without detection and perfect quantum broadcasting. 

The most widely accepted model for calculating the evolution of a quantum system in the presence of closed timelike curves, proposed by Deutsch~\cite{Deutsch91a} (see also Politzer~\cite{Politzer94a}), involves a self-consistent solution for the density matrix. In this model, a unitary interaction $U$ of a chronology respecting (CR) quantum system with a quantum system traveling around the closed timelike curve (CTC) leads to self-consistent evolution of an initial state which does not give rise to any of the typical ``patricidal paradoxes'' usually associated with time travel~\cite{Deutsch91a}. As Deutsch's solution relies only on the geometry of spacetime described by general relativistic closed timelike curves, we refer to the CTCs in our study as  ``geometric closed timelike curves''. The quantum systems are the density matrices of quantum mechanics and the dynamics are augmented from the usual linear evolution. For each input density matrix $\rho_{\text{CR}}$, the CTC quantum system is postulated to find at least one fixed-point  $\rho_{\text{CTC}}$ such that 
\begin{equation}\label{Eqn:DeutschEq1}
\rho_{\text{CTC}}=\text{Tr}_{\text{CR}}\left(U\,\rho_{\text{CR}}\otimes \rho_{\text{CTC}} \,U^{\dagger}\right),
\end{equation}
which is called a self-consistency condition for the CTC system~\cite{Deutsch91a}. The final state of the CR system is then defined in terms of the fixed point as~\cite{Deutsch91a}
\begin{equation}\label{Eqn:DeutschEq2}
\rho'_{\text{CR}}=\text{Tr}_{\text{CTC}}\left(U\,\rho_{\text{CR}}\otimes \rho_{\text{CTC}}\, U^{\dagger}\right).
\end{equation}
The induced mapping $\rho_{\text{CR}}\to\rho'_{\text{CR}}$ is nonlinear because the fixed point $\rho_{\text{CTC}}$ also depends on the input state $\rho_{\text{CR}}$. It is this nonlinearity that would distinguish the CTC system from ordinary quantum mechanics.
 
It is an interesting question from a fundamental physics point of view whether operations forbidden by the linearity of quantum mechanics would be permissible in the presence of CTC systems. Previously, it has been argued that the CTC nonliearity could improve quantum state discrimination~\cite{Brun09a} or speed up hard computations~\cite{Bacon04a}. An alternative viewpoint that such increased power is not implied by CTCs~\cite{Bennett09a} has been argued to be inconsistent with the Deutsch model~\cite{Ralph10a, Cavalcanti10a}. \textcolor{black}{
In particular we take the common view that if pure states are deterministically prepared and inserted into the circuit by some party, then their dynamics are calculated shot by shot via the Deutsch recipe.}

It was discovered by Wooters and Zurek~\cite{Wooters82a} almost three decades ago that the linearity of quantum mechanics leads to an impossibility theorem called the ``no cloning theorem''. The theorem dictates that no apparatus exists that will copy an arbitrary quantum state. It does not rule out the possibility of copying orthonormal states by a device designed especially for that purpose, but it does rule out the existence of a device capable of cloning an arbitrary state. 
\textcolor{black}{The possibility of cloning was discussed in the original Deutsch model \cite{Deutsch91a}. However an explicit protocol was not proposed.}

In this Letter, we first show that an apparatus exists that will clone a quantum state chosen randomly from a finite alphabet of states in the presence of a closed timelike curve. We then explicitly construct such a device and show it allows ideal cloning of a finite alphabet. Finally, we construct symmetrized versions of our cloner that, on average, act equally on arbitrary input states. We evaluate the fidelity of our clones with the original state and show that this fidelity exceeds the no-cloning bound for sufficiently large ancillar dimensions.


The general problem, posed formally, is as follows: A CR quantum system $AB$ is composed of two parts, $A$ and $B$, each belonging to an $N$ dimensional Hilbert space. System $A$ is prepared in one state from a set $A=\{\rho_{j}\}_{j=0}^{N-1}$ of $N$ quantum states. System $B$, slated to receive the unknown state, is in a standard quantum state $\Sigma$. The initial state of the composite CR system $AB$ is in the product state $\rho_{s}\otimes \Sigma$, where $s=0, 1, \cdots, N-1$ specifies which state is to be cloned. We ask whether there is any physical process that leads to an evolution of the form 
\begin{equation}\label{Eqn:CloneCTCEq1}
\text{Tr}_{\text{CTC}}\left(U\,\rho_{s}\otimes\Sigma\otimes \rho_{\text{CTC}}^{s} \,U^{\dagger}\right)=\rho_{s}\otimes \rho_{s},
\end{equation}
for some unitary operator $U$ and a fixed point $\rho_{\text{CTC}}^{s}$ which satisfies a self-consistency condition for the CTC system
\begin{equation}\label{Eqn:CloneCTCEq2}
\rho_{\text{CTC}}^{s}=\text{Tr}_{AB}\left(U\,\rho_{s}\otimes\Sigma\otimes \rho_{\text{CTC}}^{s} \,U^{\dagger}\right),
\end{equation}
for each $s$. To demonstrate how to circumvent the no-cloning theorem, we employ the concept of fidelity $F(\rho_{i},\rho_{j})$ between two density operators, defined by~\cite{Barnum96a, Jozsa94a} 
\begin{equation}\label{Eqn:FidDefEq}
F(\rho_{i},\rho_{j})=\text{Tr}\left(\sqrt{\sqrt{\rho_{i}}\rho_{j}\sqrt{\rho_{i}}}\right),
\end{equation}
where for any positive operator $O$,  $\sqrt{O}$ denotes its unique positive square root. Fidelity is an analog of the modulus of the inner product for pure states\footnote{\textcolor{black}{Fidelity is also commonly defined as the square of the RHS of (\ref{Eqn:FidDefEq}).
}} and can be interpreted as a measure of distinguishability for quantum states: it ranges between 0 and 1, reaching 0 if and only if the states are orthogonal and reaching 1 if and only if $\rho_{i}=\rho_{j}$. It is invariant under the interchange $i\leftrightarrow j$ and under the unitary transformation $\rho_{s}\to U\rho_{s}U^{\dagger}$ for any unitary transformation $U$~\cite{Jozsa94a}. Also, from the properties of the direct product, we have~\cite{Jozsa94a}
\begin{equation}\label{Eqn:FidPropEq1}
F(\rho_{i}\otimes\sigma_{i}, \rho_{j}\otimes\sigma_{j})=F(\rho_{i},\rho_{j})F(\sigma_{i},\sigma_{j}).
\end{equation}			
Furthermore, if $\sigma=\text{Tr}_{C}(\tilde{\sigma})$ and $\tau=\text{Tr}_{C}(\tilde{\tau})$ where $\text{Tr}_{C}$ denotes partial trace over the subsystem $C$, we have~\cite{Barnum96a, Jozsa94a} 
\begin{equation}\label{Eqn:FidPropEq2}
F(\tilde{\sigma},\tilde{\tau})\leq F(\sigma,\tau),
\end{equation}			
referred to as the partial trace property. Equality holds when there is an optimal positive operator-valued measure (POVM)~\cite{Barnum96a, Jozsa94a} . 

When there is no CTC system interacting with the CR system $AB$, then the cloning condition is simplified such that 
\begin{equation}\label{Eqn:FidPropEq3}
\text{Tr}_{C}\left(U\,\rho_{s}\otimes\Sigma\otimes Y \,U^{\dagger}\right)=\rho_{s}\otimes\rho_{s},
\end{equation}
where $C$ is an auxiliary quantum system in some standard state $Y$. In this case, it can be shown that the optimal POVM exists and from Eqs.~(\ref{Eqn:FidPropEq1}) to~(\ref{Eqn:FidPropEq3}), we obtain
\begin{align*}
F(\rho_{i},\rho_{j})&=F\left(\text{Tr}_{C}\left(U\,\rho_{i}\otimes\Sigma\otimes Y \,U^{\dagger}\right), \text{Tr}_{C}\left(U\,\rho_{j}\otimes\Sigma\otimes Y \,U^{\dagger}\right)\right)\\
&= F(\rho_{i}\otimes\rho_{i}, \rho_{j}\otimes\rho_{j})=F(\rho_{i},\rho_{j})^{2},
\end{align*}
which means that $F(\rho_{i},\rho_{j})=1$ or 0; i.e. $\rho_{i}$ and $\rho_{j}$ are identical or orthogonal. As a result, there can be no cloning for density operators with nontrivial fidelity when there is no violation of chronology~\cite{Barnum96a}. 

On the other hand, when the CR system $AB$ is interacting with the CTC system, from the properties of the direct product, we have
\begin{equation}\label{Eqn:FidPropCTCEq}
\begin{aligned}
&F\left(U\,\rho_{i}\otimes\Sigma\otimes \rho_{\text{CTC}}^{i} \,U^{\dagger}, U\,\rho_{j}\otimes\Sigma\otimes \rho_{\text{CTC}}^{j} \,U^{\dagger}\right)\\
&\quad\quad\quad\quad\quad\quad\quad\quad=F(\rho_{i}, \rho_{j})F(\rho_{\text{CTC}}^{i}, \rho_{\text{CTC}}^{j}).\end{aligned}
\end{equation}
Assuming~(\ref{Eqn:CloneCTCEq1}) and making use of~(\ref{Eqn:CloneCTCEq2}) and~(\ref{Eqn:FidPropEq2}), we have the following partial trace properties for the CTC system:
\begin{equation}\label{Eqn:PTracePropCTCEq1}
F(\rho_{i},\rho_{j})F(\rho_{\text{CTC}}^{i}, \rho_{\text{CTC}}^{j})\leq F(\rho_{i},\rho_{j})^{2},
\end{equation}
and
\begin{equation}\label{Eqn:PTracePropCTCEq2}
F(\rho_{i},\rho_{j})F(\rho_{\text{CTC}}^{i}, \rho_{\text{CTC}}^{j})\leq F(\rho_{\text{CTC}}^{i},\rho_{\text{CTC}}^{j})\,\,\text{for}\,\, i\neq j.
\end{equation}

Due to the requirement of different fixed points $\rho_{\text{CTC}}^{i}$ and $\rho_{\text{CTC}}^{j}$, the existence of an optimal POVM for equalities in Eqs.~(\ref{Eqn:PTracePropCTCEq1}) and~(\ref{Eqn:PTracePropCTCEq2}) is not guaranteed. From Eq.~(\ref{Eqn:PTracePropCTCEq1}), we have $F(\rho_{i},\rho_{j})\geq0$  or $F(\rho_{\text{CTC}}^{i},\rho_{\text{CTC}}^{j})\leq F(\rho_{i},\rho_{j})$ for non trivial fidelity for cloning of density operators when the CR quantum system is interacting with the CTC system. 

\textcolor{black}{While eqs. (10) and (11) are necessary conditions for any state specified as initial data  to be copied faithfully via
the CR system,   it is difficult to prove that these equations are also sufficient conditions. Instead, we pursue an explicit construction of a universal cloner and show numerically that it exceeds the non cloning bound for any unknown state.~\cite{Buzek96a, Scarani05a}} 

\begin{figure}[ht!]
\includegraphics[width=6.9cm]{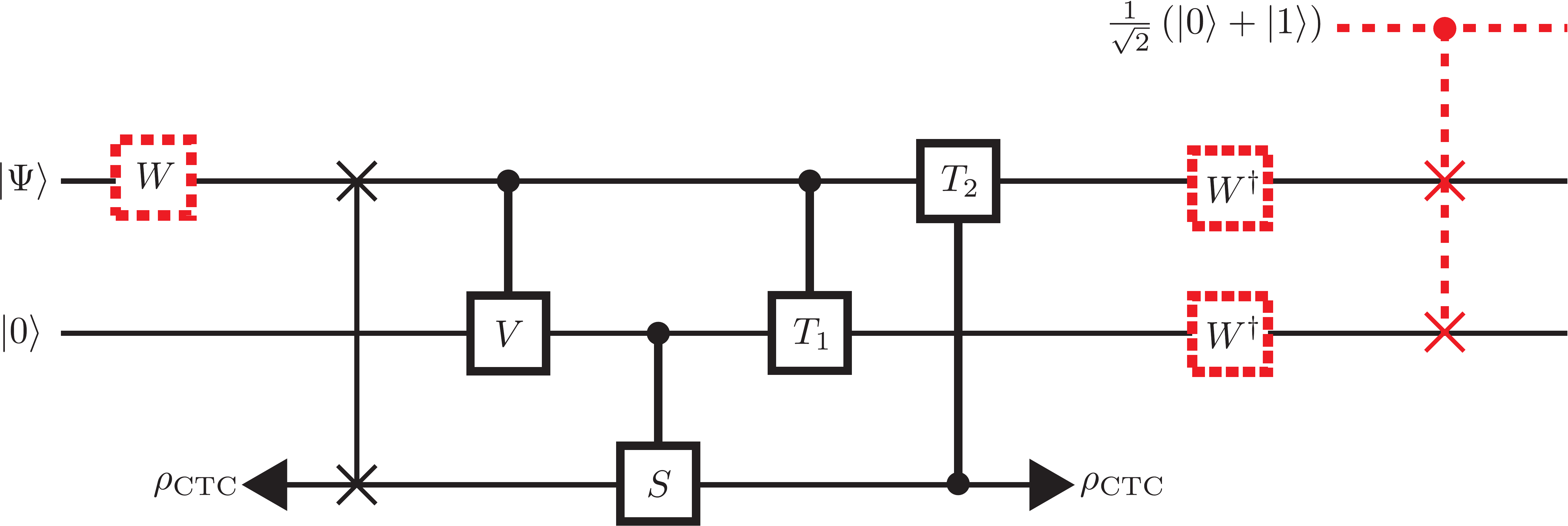}
\caption{\footnotesize (Color online) Circuit to broadcast the set of $N$ states $\{ \ket{\psi_{j}}\}^{N-1}_{j=0}$ is shown with solid lines. Circuit to clone an arbitrary qubit $\ket{\Psi}$ that is both input state independent and symmetric requires the additional dashed lined (red) components.}
\label{BroadcastFig:Fig}
\end{figure}

As an example, consider a set $\{\ket{\psi_{j}}\}_{j=0}^{N-1}$ of $N$ distinct states in a space of dimension $N$. The set $\{\ket{\psi_{j}}\}$ is not necessarily an orthonormal set. It can be shown~\cite{Brun09a} that there is a unitary transformation  $U_{j}$ such that  $U_{j}\ket{\psi_{j}}=\ket{j}$, provided $\bra{j}U_{k}\ket{\psi_{j}}\neq0$, $\forall\, j, k$, where the states $\ket{j}$ are a standard orthonormal basis for the $N$-dimensional Hilbert space. We now construct a CTC containing an $N$-dimensional system in the loop. We prepare the input system $A$ consisting of one of the states $\ket{\psi_{j}}$. The input system $B$ is prepared as $\Sigma=\kb{0}{0}$. The evolution operator $U$ for the total system in the presence of a CTC is given by $U=T_{2}T_{1}SVW$, as shown in the solid lined part of Fig.~(\ref{BroadcastFig:Fig}), where $W=\text{SWAP}(A\leftrightarrow \text{CTC})$, $V=\text{CSUM}\otimes \id_{\text{CTC}}$,  $S=\id_{A}\otimes\sum_{k}\kb{k}{k}\otimes U_{k}$, $T_{1}=\sum_{l}\kb{l}{l}\otimes U_{l}^{\dagger}\otimes\id_{\text{CTC}}$ and $T_{2}=\sum_{m}U_{m}^{\dagger}\otimes\id_{B}\otimes\kb{m}{m}$. Here, CSUM acts on an orthonormal basis according to $\text{CSUM}\left(\ket{i}\otimes\ket{j}\right)=\ket{i}\otimes\ket{j+i(\text{mod}\, N)}$~\cite{Gottesman04A}. Before the interaction, the CTC system is in the state $\rho_{\text{CTC}}$ and the chronology respecting system $AB$ is in the state $\rho_{\text{CR}}=\kb{\psi_{j}}{\psi_{j}}\otimes\Sigma$. 

It is straightforward to show that the solution $\rho_{\text{CTC}}=\kb{j}{j}$ uniquely satisfies the self-consistency condition given by Eq.~(\ref{Eqn:DeutschEq1}). The output state of the chronology respecting system is given by
\begin{equation}\label{Eqn:CloningCRoutput}
\begin{aligned}
\rho'_{\text{CR}}&=\text{Tr}_{\text{CTC}}\left(U\,\kb{\psi_{j}}{\psi_{j}}\otimes\Sigma\otimes\kb{j}{j}\,U^{\dagger}\right)\\
&=\left(U_{j}^{\dagger}\kb{j}{j}U_{j}\right)\otimes \left(U_{j}^{\dagger}\kb{j}{j}U_{j}\right)\\
&=\kb{\psi_{j}}{\psi_{j}}\otimes\kb{\psi_{j}}{\psi_{j}},
\end{aligned} 
\end{equation}
where again $U=T_{2}T_{1}SVW$, which shows that the CTC system indeed allows the cloning of arbitrary pure quantum states. It is clear that the above solution satisfies the cloning condition $F(\rho_{\text{CTC}}^{i},\rho_{\text{CTC}}^{j})\leq F(\rho_{i},\rho_{j})$ because $\rho_{\text{CTC}}^{j}$ are orthogonal and $F(\rho_{\text{CTC}}^{i},\rho_{\text{CTC}}^{j})=0$ for $i\neq j$. This is an example of perfect broadcasting in which the density operator of each of the separate systems is the same as the state to be broadcast~\cite{Barnum96a}. 


We can extend this broadcasting result to show that the solid lined part of Fig.~(\ref{BroadcastFig:Fig}) can also clone an arbitrary qubit state $\ket{\Psi}=\alpha\ket{0}+\beta\ket{1}$ with a fidelity that exceeds the no cloning bound of $\sqrt{5/6}\approx91.287\%$~\cite{Buzek96a, Scarani05a}. The ability of the broadcasting circuit to clone an arbitrary state $\ket{\Psi}$ is strongly dependent on the number of states $N$ it is set up to broadcast. To investigate this relationship we broadcast $N$ Bloch sphere surface states $\{\ket{\psi_{j}}\}_{j=0}^{N-1}$, \textcolor{black}{each of the form $\ket{\psi_{j}}=\cos(\frac{\theta_{j}}{2})\ket{0}+e^{i\phi_{j}}\sin(\frac{\theta_{j}}{2})\ket{1}$, where $0\leq\theta\leq\pi$, $0\leq\phi_{j}\leq 2\pi$}, and calculate the resulting output fidelity. It should be noted at this point that even though we are broadcasting states from the surface of the Bloch sphere, i.e. qubit states, in order to broadcast $N$ states we must map our original qubit states onto an $N$-dimensional Hilbert space. 

In order to maximise the output cloning fidelity the intuitive choice for which states on the Bloch sphere to broadcast is to have them equidistant from each other. However, choosing $N$ equidistant points on the surface of a sphere cannot be solved analytically in general~\cite{ConwayBook98}. We instead rely on numerical solutions~\cite{SloaneElectron} to the so-called Thomson problem, where we consider the equilibrium configuration of $N$ electrons on the surface of a sphere, such that the potential energy is minimised. 


Once the $N$ states to broadcast $\{\ket{\psi_{j}}\}_{j=0}^{N-1}$ have been chosen, we build the solid lined part of Fig.~(\ref{BroadcastFig:Fig}) by explicitly constructing the unitaries $U_{k}$ from the recipe given by Brun~\etal~in~\cite{Brun09a}. We assume the CTC is in the completely general state $\rho_{\text{CTC}}=\sum_{m,n=0}^{N-1}\lambda_{mn}\kb{m}{n}$. Solving the consistency relation in Eq.~(\ref{Eqn:DeutschEq1}) is equivalent to finding the $+1$ eigenvector of the matrix
\begin{equation}\label{lambdaEqn}
\sum_{\substack{   a,b \\  m,n  }=0}^{N-1}\bra{n}U_{b}U_{a}^{\dagger}\ket{m}\bk{\psi_{n}}{\psi_{m}}\bra{a}U_{m}\kb{\Psi}{\Psi}U_{n}^{\dagger}\ket{b}\kb{a,b}{m,n}.
\end{equation}
Once $\rho_{\text{CTC}}$ has been evaluated for the given state $\ket{\Psi}$, we can calculate the output fidelities for the two output modes, where we use the pure state simplification of Eq.~(\ref{Eqn:FidDefEq})
\begin{align}
F_{1}&=\Bigl[\sum_{i,m,n}\lambda_{mn}\bra{i}U_{m}\ket{\Psi}\bra{\Psi}U_{n}^{\dagger}\ket{i} \bk{\psi_{n}}{\psi_{m}}\bra{\Psi}U_{i}^{\dagger}\kb{m}{n}U_{i}\ket{\Psi}\Bigr]^{\frac{1}{2}},\nonumber\\
F_{2}&=\Bigl[\sum_{i,n}\lambda_{nn}\bra{i}U_{n}\ket{\Psi}\bra{\Psi}U_{n}^{\dagger}\ket{i} \bk{\Psi}{\psi_{n}}\bk{\psi_{n}}{\Psi}\Bigr]^{\frac{1}{2}}.\label{OutputFidEqn}
\end{align}

\begin{figure}[ht!]
\includegraphics[width=7.9cm]{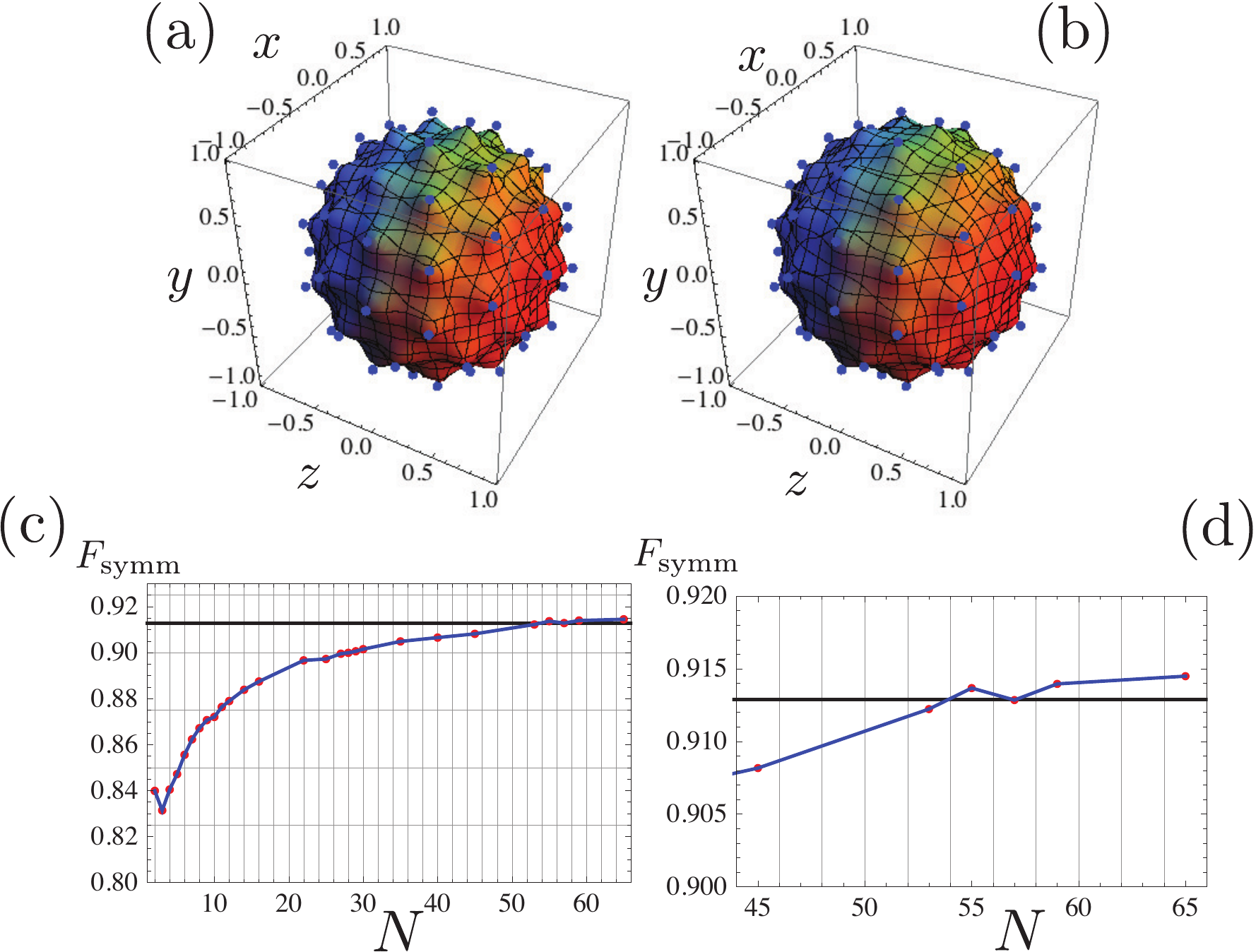}
\caption{\footnotesize (Color online) In (a) and (b) the raw fidelity data $F_{1}$ and $F_{2}$ are shown, respectively, for the 10,000 random points chosen to clone when we broadcast $N=65$ points with the solid lined portion of Fig.~(\ref{BroadcastFig:Fig}). The small blue points show the 65 broadcast states. In (c) we plot the resulting symmetric fidelity as a function of $N$ when all the components in Fig.~(\ref{BroadcastFig:Fig}) are included. (d) is a magnification of (c). The solid black line corresponds to $F=\sqrt{5/6}\approx 91.287\%$.} 
\label{AvAvFidPlotPaperN65:Fig}
\end{figure}

We examine the ability of the broadcasting circuit in Fig.~(\ref{BroadcastFig:Fig}) to clone by setting $\ket{\Psi}$ to be a random point on the surface of the bloch sphere and calculate the fidelities from Eq.~(\ref{OutputFidEqn}). We repeat this for 10,000 random points on the Bloch sphere and calculate the average fidelity. In Figs.~(\ref{AvAvFidPlotPaperN65:Fig}a) and~(\ref{AvAvFidPlotPaperN65:Fig}b), we show the raw data for the $N=65$ case. The two output fidelities $F_{1}$ and $F_{2}$ are not symmetric, as would be expected from the asymmetric form of Eq.~(\ref{OutputFidEqn}). Also, due to the CTC interaction, the the output fidelities are state dependent. That is, if we choose a point on the Bloch sphere to be cloned that is close to one of the $N$ broadcast states $\{\ket{\psi}_{j}\}_{j=0}^{N-1}$, we expect it to be cloned with a high fidelity, whereas if the point is far from any $\ket{\psi_{j}}$, we would expect a low fidelity. 
To obtain a quantum state cloner that is symmetric and state independent, we include the additional dashed components in Fig.~(\ref{BroadcastFig:Fig}), where the random unitary $W$ acts to make the device state independent and the controlled swap gate leads to symmetric output states. The result of this state independent, symmetric circuit can be seen in Fig.~(\ref{AvAvFidPlotPaperN65:Fig}c) and~(\ref{AvAvFidPlotPaperN65:Fig}d), where we plot the symmetric output fidelity as a function of $N$. As can be seen, we expect the circuit shown in Fig.~(\ref{BroadcastFig:Fig}) to break the quantum cloning bound when we choose to broadcast at least 55 states from the surface of the Bloch sphere.  Note that there is some freedom when defining the $U_{k}$ for the set $\{\ket{\psi_{k}}\}_{j=0}^{N-1}$ according to the Brun~\etal~recipe~\cite{Brun09a}. This means that the fidelities in Figs.~(\ref{AvAvFidPlotPaperN65:Fig}c) and~(\ref{AvAvFidPlotPaperN65:Fig}d) may be improved with a different choice of $U_{k}$.


One of the original arguments against cloning was that it would allow signalling, i.e. faster than light communication, when applied to an entangled state. Let's assume that one party of the entangled state, say Alice is locally interacting with the CTC for the cloning.  For example, if the state vector of the entangled state is given by	
\begin{equation*}
\ket{\Psi}_{AR}=\frac{1}{\sqrt{2}}\left(\ket{0}_{A}\ket{1}_{R}+\ket{1}_{A}\ket{0}_{R}\right),
\end{equation*}
then the output of the chronology respecting system would be
\begin{equation}
\rho_{\text{tot}}=\text{Tr}_{\text{CTC}}\left(U\,\ket{\Psi}_{AR}\bra{\Psi}\otimes\kb{\Sigma}{\Sigma}\otimes\rho_{\text{CTC}}\,U^{\dagger}\right).
\end{equation}
Here the unitary operator $U$ is not acting on the state belonging to Rob and takes the form $U=VW_{1}W_{2}$ where $W_{1}=\text{SWAP}\left(A\leftrightarrow \text{CTC}\right)$, $W_{2}=\id_{A}\otimes \text{SWAP}\left(B\leftrightarrow \text{CTC}\right)$ and $V=\id_{A}\otimes\text{CSUM}$. By taking the partial trace with respect to Rob's state and from Eqs.~(\ref{Eqn:CloneCTCEq1}) and~(\ref{Eqn:CloningCRoutput}), we get 
\begin{align}
\text{Tr}_{R}\left(\rho_{\text{tot}}\right)&=\text{Tr}_{\text{CTC}}\left(U\,\text{Tr}_{R}\left(\ket{\Psi}_{AR}\bra{\Psi}\right)\otimes \kb{\Sigma}{\Sigma}\otimes\rho_{\text{CTC}}\,U^{\dagger}\right) \nonumber\\
&=\text{Tr}_{\text{CTC}}\left(U\,\rho_{A}\otimes\kb{\Sigma}{\Sigma}\otimes\rho_{\text{CTC}}\,U^{\dagger}\right)\label{Eqn:Signalling}\\
&=\rho_{A}\otimes\rho_{A}\nonumber,
\end{align}
where $\rho_{A}=\text{Tr}_{R}\left(\ket{\Psi}_{AR}\bra{\Psi}\right)$. 

Since the clone is the reduced density operator of the initial entangled state $\ket{\Psi}_{AB}$, no correlations remain between the clone and the other half of the entangled state. This clearly denotes that faster than light communication does not result from this type of cloning. 

Our results have further implications. In quantum cryptography, the legitimate users of a communication channel encode the bits 0 and 1 into nonorthogonal pure states to ensure that any eavesdropping is detectable since eavesdropping necessarily disturbs the state sent to the legitimate user due to the no-cloning theorem. If nature allows CTC's, an eavesdropping party with access to a CTC can prepare the ancillary state $\Sigma$ and obtain a perfect copy of the input state initially possessed by the system $A$. However, entanglement-based QKD (quantum key distribution) would remain secure against the type of cloning we described in this work because there is no correlation as shown by Eq.~(\ref{Eqn:Signalling}). 

\textcolor{black}{In conclusion, we have shown that we can violate the fidelity no-cloning bound for completely arbitrary qubit states, provided we have access to a CTC. We expect the fidelity curve in Fig.~(\ref{AvAvFidPlotPaperN65:Fig}c) to continue towards 1 as N increases above N=65. However, we do not at present have the numerical resources to test this claim.} From an historical point of view, many insights obtained from the analysis of thought experiments that might be impossible to actually realize contributed significantly to the development of quantum mechanics~\cite{Deutsch91a}. Investigations of quantum mechanics in the presence of the closed timelike curves, even if they remain only theoretical constructs, may well contribute to the development of a yet unknown full theory of quantum gravity.

\begin{acknowledgments}

D.~A. was supported by the research grant from Korea Commications Agency through Contract No. KCA-12-911-04-003. R.~B.~M. was supported by the Natural Sciences and Engineering Research Council of Canada.

\end{acknowledgments}


\end{document}